\documentclass[prb,twocolumn,superscriptaddress,showpacs,showkeys]{revtex4}
\usepackage{graphics}% Include figure files
\usepackage{color}
\usepackage{amsmath}% for advanced Eq. numering
\usepackage{amssymb}% for advanced Eq. numering
\usepackage[colorlinks=false]{hyperref}
\usepackage{pslatex}
\usepackage{amsfonts}
\usepackage{amsthm}
\usepackage{graphicx}

\include{MyMc} %simple macros allowed
\newcommand{\dFzpi}{\ensuremath{d_F^\mathrm{0\text{-}\pi}}\xspace}

\begin{document}

\title{
  Josephson junctions with negative second harmonic in the current-phase relation: properties of novel $\varphi$-junctions
}

\author{E.~Goldobin}
\email{gold@uni-tuebingen.de}
\author{D.~Koelle}
\author{R.~Kleiner}
\affiliation{%
  Physikalisches Institut II --- Experimentalphysik II,
  Universit\"at T\"ubingen,
  Auf der Morgenstelle 14,
  D-72076 T\"ubingen, Germany
}

\author{A.~Buzdin}
\affiliation{%
  Institut Universitaire de France and University Bordeaux I, 
  CPMOH, UMR 5798, 33405 Talence, France
}

\pacs{
  74.50.+r,   %Proximity effects, weak links, tunneling phenomena,
              %and Josephson effect
  85.25.Cp    %Josephson devices
  74.20.Rp    %Pairing symmetries (other than s-wave)
}

\keywords{
  Long Josephson junction, sine-Gordon,
  half-integer flux quantum, semifluxon,
  0-pi-junction
}

%Acronims:
% LJJ:66, SF, PSF, NSF, AFM:20, ZFS:9

%\definecolor{gray}{gray}{0.75}
%\date{9.9.2002 [\colorbox{gray}{cond-mat/0209214}]}
\date{\today}

\begin{abstract}

Several recent experiments revealed a change of the sign of the first harmonic in the current-phase relation of Josephson junctions (JJs) based on novel superconductors, \eg, $d$-wave based JJs or JJ with ferromagnetic barrier. In this situation the role of the second harmonic can become dominant; in this case, it determines the scenario of a 0-$\pi$ transition. We discuss different mechanisms of the second harmonic generation and its sign. If the second harmonic is negative, the 0-$\pi$ transition becomes continuous and the realization of a so-called $\varphi$ junction is possible. We study the unusual properties of such a novel JJ such as critical currents, magnetic field penetration, plasma gap, and microwave response. We also analyze the possible experimental techniques for their observation.

\end{abstract}

%{\small
%  To be submitted to Phys.\ Rev.\ B (many pages)
%}

\maketitle

\section{Introduction}
\label{Sec:Intro}

Several novel Josephson junctions (JJs), \eg, $s$-wave/$d$-wave JJs\cite{Smilde:ZigzagPRL,Ariando:Zigzag:NCCO,Hilgenkamp:zigzag:SF}, $d$-wave/$d$-wave JJs\cite{Tsuei:Review,Mints:2002:SplinteredVortices@GB}, superconductor-ferromagnet-superconductor (SFS)\cite{Ryazanov:2001:SFS-PiJJ,Blum:2002:IcOscillations,Bauer:2004:SFS-SpontSuperCurrents,Sellier:2004:SFS:HalfIntShapiro,Oboznov:2006:SFS-Ic(dF)} or superconductor-insulator-ferromagnet-superconductor (SIFS) JJs\cite{Kontos:2002:SIFS-PiJJ,Weides:2006:SIFS-HiJcPiJJ}, etc.\cite{Vavra:2006:SIfIS} can have a current phase relation
\begin{equation}
  j_s = j_{c1}\sin(\phi)
  , \label{Eq:CPR1}
\end{equation}
with the critical current density $j_{c1}$ changing its sign, e.g., as a function of temperature $T$\cite{Ryazanov:2001:SFS-PiJJ,Blum:2002:IcOscillations,Weides:2006:SIFS-HiJcPiJJ} or F-layer thickness\cite{Oboznov:2006:SFS-Ic(dF),Kontos:2002:SIFS-PiJJ,Weides:2006:SIFS-HiJcPiJJ}. A negative sign of $j_{c1}$ leads to the formation of a so-called $\pi$ junction with a ground state phase $\phi=\pi$\cite{Golubov:2004:CPR,Buzdin:2005:Review:SF}. Usually, the single harmonic current-phase relation (CPR) is adequate for the description of the JJ properties; high order harmonic terms in $j_s(\phi)$ can be omitted. However at the 0-$\pi$ transition $j_{c1}$ vanishes and the second harmonic becomes important. Then, one can write
\begin{equation}
  j_s(\phi) = j_{c1}\sin(\phi) + j_{c2}\sin(2\phi)
  , \label{Eq:CPR}
\end{equation}
see Ref.~\onlinecite{Golubov:2004:CPR} and references therein.

A second harmonic with negative amplitude $j_{c2}$ may result in the formation of a so-called $\varphi$-junction, \ie, a junction for which, in the absence of a supercurrent, $j_s=0$, the phase drop across the junction $\phi=\pm\varphi$ such that $0<\varphi<\pi$, see Ref.~\onlinecite{Buzdin:2003:phi-LJJ}.

In this paper, we investigate the rich physics of JJs with a substantial second harmonic in the CPR, especially when $j_{c2}$ is negative. We discuss short as well as long JJs and propose several experimental tests of our predictions.

The paper is organized as follows. In Sec.~\ref{Sec:Origin} we discuss the origin of the second harmonic in the CPR for several types of JJs. In the main Sec.~\ref{Sec:Results} we introduce the model, derive conditions for the existence of a $\varphi$ JJs and discuss their properties such as critical currents, magnetic field penetration, plasma gap, and microwave response. Sec.~\ref{Sec:Conclusions} concludes this work.

\iffalse
argue that the origin of appearance of the second harmonic in the current phase relation is intimately related with the relative sign of the first and the second harmonic. The Mints's \emph{faceting mechanism always results in the negative second harmonics} while all existing theories of $a$-$b$-plane Josephson effect predict positive second harmonic. We also discuss the transition from 0 to $\pi$ state and predict unusual properties of JJ around this transition.
\fi

\section{Origin of the second harmonic}
\label{Sec:Origin}

\iffalse
%
\begin{table}
  \begin{tabular}{ccc}
    system & $I_{c2}$ & Ref.\\
    \hline
    Asymm. $45^\circ$ GB JJ & $I_{c2}<0$ & \cite{Ilichev:2001:Sym45GBJJ:DegenGrdStt}\\
    \hline
    e & f & g\\
    \hline
    e & f & g\\
    \hline
  \end{tabular}
  \label{Tab:}
\end{table}
%
\fi

\subsection{d-wave superconductors}
The second harmonic in the CPR was observed exprimentally in symmetric\cite{Ilichev:1998:Sym45GBJJ:NonSinCPR,Ilichev:2001:Sym45GBJJ:DegenGrdStt} $45^\circ$ [001] tilt grain boundary (GB) JJs, in tilt-twist GB JJs\cite{Bauch:2007:YBCO-45GBJJ:MQT} and in $c$-axis $s$-wave/$d$-wave \YBCO-Nb JJs\cite{Komissinski:2002:Nb-cYBCO:CPR:2ndHarm}.
The possibility that the first harmonic vanishes at some temperature was predicted theoretically\cite{Kashiwaya:2000:SurfBoundStt} and was observed in experiment\cite{Ilichev:2001:Sym45GBJJ:DegenGrdStt,Testa:2001}.

Calculations show that the intrinsic second harmonic term for bicrystal $45^\circ$ grain-boundary JJs made of $d$-wave superconductors is negative\cite{Kashiwaya:2000:SurfBoundStt}. For more details turn to reviews.\cite{Golubov:2004:CPR,Tafuri:2005:HTS:WeakLinks}

\paragraph*{Faceting.} Even in the cases, when the second harmonic was observed experimentally, it was not clear whether it is present intrinsically or it is a result of faceting or interface roughness. In the latter case, the second harmonic appears in the equation for the average (slowly varying) phase, which changes on a length scale much larger than the interface roughness. The theory of this ``effective negative second harmonic'' was developed by R. Mints and coauthors\cite{Mints:1998:SelfGenFlux@AltJc,Mints:2000:SelfGenFlux@GB,Mints:2001:FracVortices@GB,Mints:2002:NonLocal+FracVortices} and was also recently discussed in Ref.~\cite{Buzdin:2003:phi-LJJ}. In many cases the amplitude of the effective second harmonic $|j_{c2}|$ may be comparable to or even larger than the amplitude of the first harmonic $|j_{c1}|$. Bicrystal $45^\circ$ GB JJs made of $d$-wave superconductors are one example of such a system where a $\varphi$-junction may appear. Here, the so-called splintered vortices were observed\cite{Mints:2002:SplinteredVortices@GB}.

\subsection{SFS or SIFS junctions}

Next, we consider SFS or SIFS junctions. Often, their CPR is sinusoidal only near $T_c$\cite{deGennes:SuperCondM&A}. At low temperatures the higher harmonic terms become more and more important. The calculations of the CPR in SFS JJs in the \emph{clean limit} indeed reveal a strongly nonsinusoidal $j_s(\phi)$ dependence\cite{Buzdin:1982,Radovic:2001:0&pi-JJ,Chtchelkatchev:2001:SFS:0-pi-trans}. In the \emph{dirty limit}, in which most experiments are done up to now, the $j_s(\phi)$ dependence becomes almost sinusoidal when the F-layer thickness $d_{F}$ exceeds the decay length of the order parameter $\xi_{F1}$, namely\cite{Buzdin:2005:0-pi-trans}
\begin{subequations}
  \begin{eqnarray}
    j_{c1}&\sim& \exp \left( -\frac{d_F}{\xi_{F1}}\right)
    \cos\left( \frac{d_F}{\xi_{F2}} \right)
    , \label{Eq:SFS,j_c1}\\
    j_{c2}&\sim& \exp \left( -\frac{2d_F}{\xi_{F1}}\right)
    . \label{Eq:SFS,j_c2}
  \end{eqnarray}
  \label{Eq:SFS:j_c12}
\end{subequations}
In the absence of spin-flip scattering (which is rather unrealistic) in the F-layer of thickness $d_F$, the decay and oscillation lengths $\xi_{F1}$ and $\xi_{F2}$ are equal to the characteristic length $\xi_{F}$. In the presence of spin-flip scattering\cite{Buzdin:2005:0-pi-trans} $\xi_{F1}<\xi_{F2}$, but such that $\xi_{F1}\xi_{F2}=\xi_{F}^2$. Thus, in practice, for $d_F>\xi_{F1}$, the contribution of the second (and higher) harmonics may be neglected everywhere, except for the vicinity of the $0$-$\pi$ transition. At the transition point $d_F^{0\text{-}\pi}=\frac\pi2\xi_{F2}\gg\xi_{F1}$, the first harmonic term vanishes and the properties of the SFS junctions are determined by the second harmonic term\cite{Buzdin:2005:0-pi-trans}. A recent experimental study \cite{Oboznov:2006:SFS-Ic(dF)}, which demonstrates two $0\to\pi$ and $\pi\to0$ transitions on the $I_c(d_F)$ dependence in a SFS JJ with a Cu$_{0.52}$Ni$_{0.48}$ alloy as an F layer, allows to estimate $\xi_{F1}\approx 1.3\units{nm}$, while the $0\to\pi$ crossover thickness  $\dFzpi\approx 11\units{nm}$. This results in a very small value of the critical current at the $0$-$\pi$ transition, c.f. Eq.~(\ref{Eq:SFS:j_c12}), $I_{c}=I_{c2}\sim \exp(-\dFzpi/\xi_{F1})I_{c0}\sim 10^{-4}I_{c0}$, where $I_{c0}\sim\exp(-\dFzpi/\xi_{F1})$ is the critical current (dominated by the first harmonic) away from the transition, i.e. if there would be only the decaying part of the first harmonic without oscillations. Therefore, the measured non zero values for the critical current at the $0$-$\pi $ transition (see Refs.~\onlinecite{Sellier:2004:SFS:HalfIntShapiro} and \onlinecite{Frolov:2006:SFS-0-pi}) can hardly be explained by an intrinsic second harmonic contribution. The intrinsic second harmonic term in SFS JJs at the 0-$\pi$ transition is positive\cite{Buzdin:2005:0-pi-trans}. 

\paragraph*{Inhomogeneous F-layer.}
Another possible mechanism of a \emph{negative} second harmonic generation in SFS JJs is the inhomogeneity of the F layer thickness near the $0$-$\pi$ transitions. As a result, the JJ consists of alternating $0$ and $\pi$ mini-junctions, which is similar to the case of faceted GB JJs. The CPR in a diffusive SFS junction is described with good accuracy by an effective (slowly varying) phase $\psi$
\begin{equation}
  j(\psi) = 
  j_{c1}\sin(\psi)+j_{c2}\sin(2\psi)
  . \label{I1I2}
\end{equation}
For a long JJ (LJJ) with alternating current density ($j_{ca}$ and $j_{cb}$ within regions of lengths $a$ and $b$ respectively) one finds
\begin{subequations}
  \begin{eqnarray}
    j_{c1} &=&\left( aj_{ca}+bj_{cb}\right) /\left( a+b\right) , \\
    j_{c2} &=&-\frac{1}{\lambda _{Ja}^{2}}\frac{a^{2}b^{2}
    \left(j_{ca}-j_{cb}\right) ^{2}}{24\left\vert j_{ca}\right\vert \left( a+b\right) ^{2}},
  \end{eqnarray}
  \label{Eq:j_c12(j_cab)}
\end{subequations}
where 
\begin{equation}
  \lambda_{Ja}=\sqrt{\frac{\Phi_0}{2\pi\mu_0 d' |j_{ca}|}}
  , \label{Eq:lambda_Ja}
\end{equation}
and $\mu_0 d' \approx \mu_0 (d_I+2\lambda_{L})$ is the inductance (per square) of the superconducting electrodes forming the JJ, see Refs.~\onlinecite{deGennes:SuperCondM&A} and \onlinecite{Buzdin:2003:phi-LJJ} for more details. The expressions (\ref{Eq:j_c12(j_cab)}) are valid when $a,\,b\ll\lambda_{Ja}$. When $|j_{c1}/2j_{c2}|>1$ the ground state of the system has a uniform phase ($0$ or $\pi$), see below. 

However, in the presence of an applied current a spatial modulation of the phase appears with the amplitude
\begin{equation}
  \Delta \phi = %\frac{2\pi\mu_0d'}{\Phi_0}
  \frac{1}{\lambda_{Ja}^2}
  \frac{ab\left(j_{ca}-j_{cb}\right) }{32|j_{ca}|}\sin (\psi)
  \ll 1.
\end{equation}
This corresponds to the appearance of two types of fractional Josephson vortices that are soliton solutions of the double sine-Gordon equation\cite{Bullough:1980:dsG,Hudak:1981:dsG:IntModes,Kivshar:1994:RapidPert}. Such vortices were observed by SQUID microscopy in $45^\circ$ grain boundary JJs\cite{Mints:2002:SplinteredVortices@GB}.

\subsection{Experimental techniques}

Let us discuss typical experimental methods and their ability to distinguish the sign of the second harmonic.
\begin{enumerate}%{
  \item The most direct technique is a measurement of the CPR by embedding the investigated JJ in a SQUID loop\cite{Ilichev:2001:Sym45GBJJ:DegenGrdStt,Komissinski:2002:Nb-cYBCO:CPR:2ndHarm,Ilichev:1998:Sym45GBJJ:NonSinCPR}. The difficulty here is that not all types of junctions can be incorporated into a SQUID with proper $\beta_L<1$ values, where $\beta_L=2I_cL/\Phi_0$ is the is the inductance parameter of the SQUID, and $I_c$ is the total critical current of the JJ.

  \item Measurements of $I_c$ \vs magnetic field may reveal a twice shorter  period of oscillations, but the sign of the second harmonic is difficult to determine. In fact, our simulations show that $I_c(H)$ curves for LJJs with strong positive or negative second harmonics of the same amplitude look the same \emph{for short JJ} and qualitatively the same for LJJ.

  \item Measurements of sub-harmonic Shapiro steps are also quite unreliable to determine the sign of the second harmonic, as the situation is very similar to the previous case --- subharmonic steps appear for both positive and negative $j_{c2}$ and look qualitatively the same. Moreover in JJs with nonvanishing capacitance (Stewart-McCumber parameter $\beta_c=2\pi I_c R^2 C/\Phi_0>0$) fractional Shapiro steps appear in any case. Nevertheless, it seems possible to extract useful information about the second harmonic from the Shapiro step modulation even in the presence of a finite capacitance\cite{Kislinskii:2005:YBCO-Nb}.

  \item The presence of a $\varphi$-junction is a strong evidence for a negative second harmonic. In a LJJ this can be manifested through the existence of (splintered) fractional Josephson vortices of two different kinds\cite{Mints:1998:SelfGenFlux@AltJc,Mints:2000:SelfGenFlux@GB,Mints:2001:FracVortices@GB,Mints:2002:NonLocal+FracVortices} that are solitons of a double sine-Gordon equation\cite{Bullough:1980:dsG}. The existence or motion of such vortices may be detected experimentally using SQUID microscopy\cite{Mints:2002:SplinteredVortices@GB}.
  
\end{enumerate}%}

\section{Results}
\label{Sec:Results}

The double sine-Gordon equation which describes the dynamics of the Josephson phase in the LJJ in question is
\begin{equation}
  \phi_{xx}-\phi_{tt}-[\sin\phi+g\sin(2\phi)] = \alpha\phi_t-\gamma
  , \label{Eq:sG}
\end{equation}
where $g(T)=j_{c2}/j_{c1}$ is a relative amplitude of the second harmonic, which, generally speaking, is a function of the temperature $T$. Subscripts $x$ and $t$ denote partial derivatives with respect to coordinate and time, respectively. The coordinates are normalized to 
\begin{equation}
  \lambda_{J1}=\sqrt{
    \frac{\Phi_0}{2\pi \mu_0 |j_{c1}| d'}
  }
  , \label{Eq:lambda_J1}
\end{equation}
where \cite{Likharev:JJ&Circuits}
\begin{equation}
  d'=d_I
  +\lambda_1\coth\left( \frac{d_1}{\lambda_1} \right)
  +\lambda_2\coth\left( \frac{d_2}{\lambda_2} \right)
  , \label{Eq:d'}
\end{equation}
and $d_{1,2}$, $\lambda_{1,2}$ are the thicknesses and London penetration depths of the superconducting electrodes and $d_I$ is the thickness of the (insulating) barrier. The time is normalized to the inverse plasma frequency $\omega_{p1}^{-1}$, where
\begin{equation}
  \omega_{p1} = \frac{j_{c1}\Phi_0}{2\pi C}
  . \label{Eq:omega_p1}
\end{equation}
The parameter $\alpha=1/\sqrt{\beta_c}$ is the dimensionless damping parameter, and $\gamma=j/j_{c1}$ is the normalized applied bias current density assumed to be uniform. Further, $\phi(x,t)$ describes either the real phase when the second harmonic is present intrinsically or the phase averaged over facets for the case when the second harmonic appears due to faceting.

The Josephson energy density (per unit of LJJ length) is given by 
\begin{equation}
  U(\phi) = \epsilon_J \sgn(j_{c1}) w 
  \left\{ 1-\cos(\phi) + \frac{g}{2}\left[ 1-\cos(2\phi) \right] \right\}
  , \label{Eq:U(phi)}
\end{equation}
where $\epsilon_J=\Phi_0|j_{c1}|/2\pi$ sets the characteristic scale of energy density, and $w$ is the JJ width. $U(\phi)$ is defined with accuracy of a constant and this constant in Eq.~(\ref{Eq:U(phi)}) is chosen so that $U(0)=0$. This is a natural choice for conventional JJs with $j_{c1}>0$ and $j_{c2}=0$, as it corresponds to the energy minimum $U=0$ which is reached at $\phi=0$. In our more general case, $\phi=0$ does not necessarily correspond to the energy minimum, but we still will use the same reference level for $U(\phi)$ to avoid confusion.

\subsection{Ground states and $\varphi$-junction}
\label{Sec:varphi-JJ}

Consider a uniform ground state of the system, $\phi(x)=const$. In this case the analysis for the LJJ reduces to the analysis of the ground state in a point-like JJ. Let us investigate conditions at which one can obtain a $\varphi$-junction\cite{Buzdin:2003:phi-LJJ}, \ie, the junction for which
\begin{equation}
  j_s(\phi)=0\mbox{ for } \phi(x)=\const
  . \label{Eq:js0}
\end{equation}
Substituting expression (\ref{Eq:CPR}) into Eq.~(\ref{Eq:js0}) we arrive at three posible solutions 
\begin{subequations}
  \begin{eqnarray}
    \phi&=&0
    ;\label{Eq:sol:0}\\
    \phi&=&\pi
    \label{Eq:sol:pi};\\
    \phi&=&\pm\varphi
    , \label{Eq:sol:varphi}
  \end{eqnarray}
  \label{Eq:phi_0}
\end{subequations}
where 
\begin{equation}
  \varphi 
  = \arccos\left( - \frac{j_{c1}}{2j_{c2}} \right)
  = \arccos\left( - \frac{1}{2g} \right)
  . \label{Eq:varphi}
\end{equation}
The ground state (\ref{Eq:sol:varphi}) corresponds to the $\varphi$-junction\cite{Buzdin:2003:phi-LJJ}. The stable solution should correspond to the energy minimum, i.e.
\begin{equation}
  \fracd[2]{U(\phi)}{\phi}=\frac{\Phi_0}{2\pi}\fracd{j_s(\phi)}{\phi} >0
  . \label{Eq:PhiJJStabCond}
\end{equation}
Substituting each of the solution (\ref{Eq:phi_0}) into (\ref{Eq:PhiJJStabCond}) we obtain the stability conditions
\begin{subequations}
  \begin{eqnarray}
    j_{c1}&>&-2j_{c2}
    ; \label{Eq:Stab:0}\\
    j_{c1}&<&+2j_{c2}
    ; \label{Eq:Stab:pi}\\
    \frac{j_{c1}^2}{2j_{c2}} & > & 2j_{c2}
    . \label{Eq:Stab:varphi}
  \end{eqnarray}
\end{subequations}
Note that in addition to stability condition (\ref{Eq:Stab:varphi}) we should impose a condition to the argument of the $\arccos$ in Eq.~(\ref{Eq:varphi}), i.e. 
\begin{eqnarray}
  \left|\frac{j_{c1}}{2j_{c2}}\right|&\leq& 1
  . \label{Eq:cond}
\end{eqnarray}
An analysis of conditions (\ref{Eq:Stab:varphi}) and (\ref{Eq:cond}) shows that the $\varphi$ ground state may be realized only for $2j_{c2}<-|j_{c1}|$. This means that a $\varphi$-junction can be obtained from a $0$ or a $\pi$ junction with a strong \emph{negative} second harmonic. 

\begin{figure}[!htb]
  \includegraphics{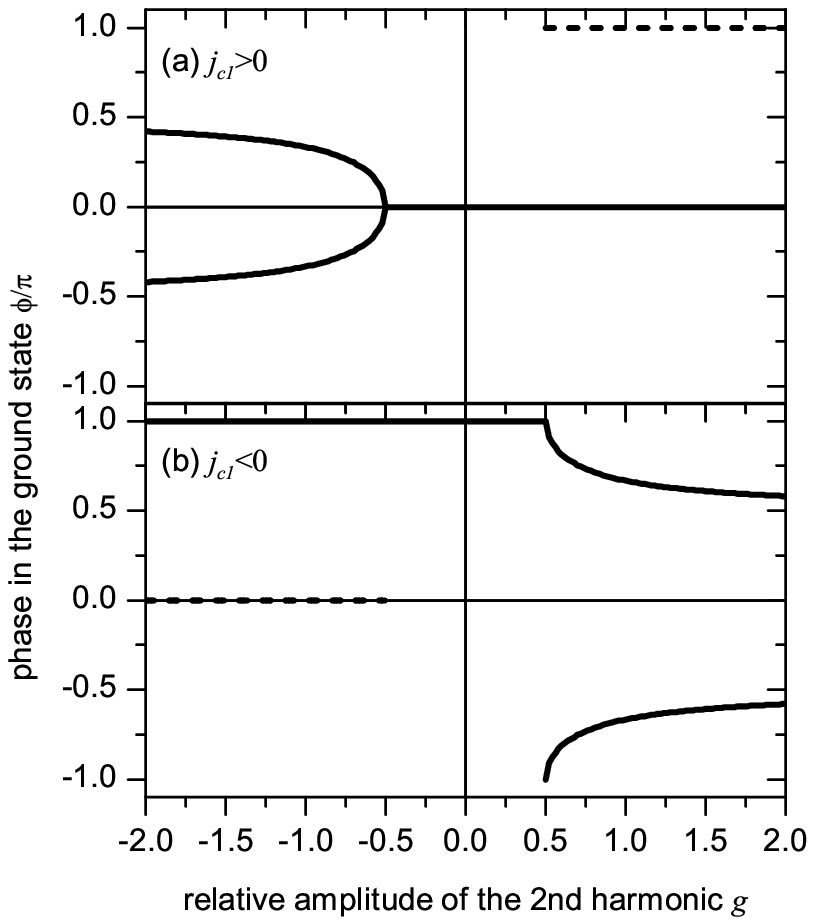}
  \includegraphics{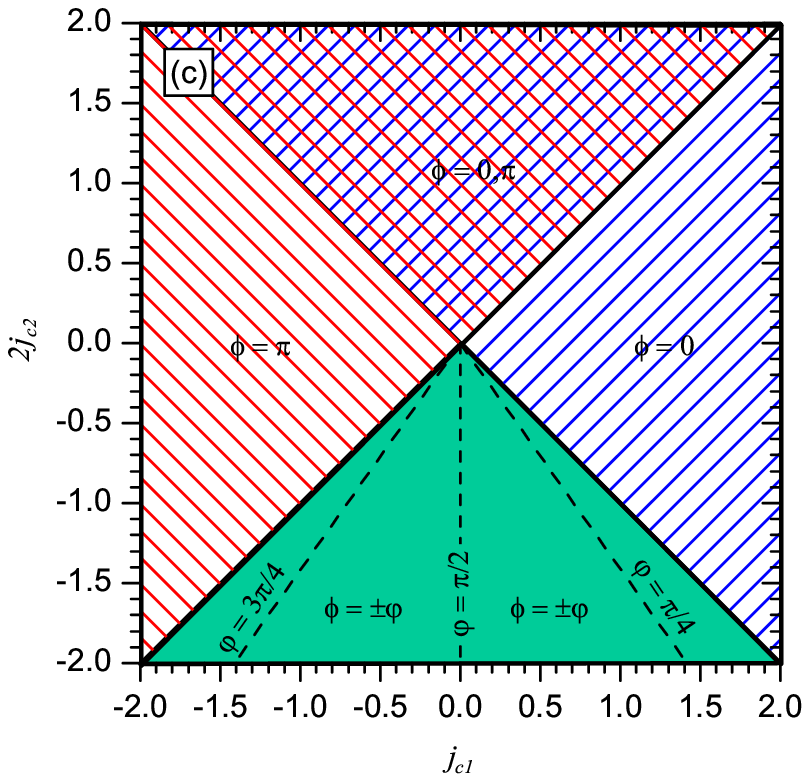}
  \caption{(Color online)
    Ground state phase as a function of $g$ for (a) $j_{c1}>0$ and (b) $j_{c1}<0$, dashed lines indicate stable states with higher energy. (c) show the ground state of the system on the plain $j_{c1}$-$j_{c2}$. 
    %(d) is the 3D representation of the ground states in $j_{c1}$,$j_{c2}$, $\phi$ space.
  }
  \label{Fig:GroundStates}
\end{figure}

Fig.~\ref{Fig:GroundStates} shows the stable uniform ground states as a function of $g$ for $j_{c1}>0$ (a) and $j_{c1}<0$ (b). The ground state diagram depends very much on the sign of $j_{c1}$. Fig.~\ref{Fig:GroundStates}c shows the regions of $0$, $\pi$ and $\pm\varphi$ ground states on $j_{c1}$-$j_{c2}$ plane.

1. Consider the case $j_{c1}>0$, see Fig.~\ref{Fig:GroundStates}a. The state $\phi=0$ is the ground state of the system for $g>-1/2$. For $g<-1/2$, there are two degenerate ground states $\phi=\pm\varphi$, see Eq.~(\ref{Eq:varphi}). In addition, for $g > 1/2$ there is a stable state $\phi=\pi$ corresponding to a local minimum of energy, but its energy is larger than the energy of the ground state $\phi=0$.

2. The case $j_{c1}<0$ ($\pi$ JJ) is shown in Fig.~\ref{Fig:GroundStates}b. The state $\phi=\pi$ is the ground state of the system for $g<1/2$. For $g>1/2$, there are two degenerate ground states $\phi=\pm\varphi$, see Eq.~(\ref{Eq:varphi}). In addition, for $g < -1/2$ there is a stable state $\phi=0$ corresponding to a local minimum of energy, but its energy is larger than the energy of the ground state $\phi=\pi$.

One may notice that the ground state diagrams shown in Figs.~\ref{Fig:GroundStates}a and b are very similar. In fact, one can reduce one case to the other by a simple transformation: $\phi\to\pi-\phi$, $g\to-g$. In fact, Eq.~(\ref{Eq:sG}) is invariant with respect to this transformation.  In this way, a 0 JJ turns into a $\pi$ JJ, and the bifurcation point turns from $g=\mp1/2$ to $\pm1/2$. Applying this transformation twice we go back to the initial case.

Below, without loosing generality, we consider only the case $j_{c1}>0$. The results for the case $j_{c1}<0$ can be naturally obtained by employing the above mentioned transformation.

%We note, that one cannot obtain a $\varphi$-junction from a $\pi$-junction by adding a positive second harmonic. In this case the transition from 0 to $\pi$ junction is discontinuous.

\subsection{Critical current}

When a JJ has two harmonics in the CPR, the natural question which arises is ``how the measured value of maximum supercurrent (critical current) $I_c$ is related to the amplitude of both harmonics $I_{c1}$ and $I_{c2}$?'' 
To answer this question we rewrite Eq.~(\ref{Eq:CPR}) in normalized units
\begin{equation}
  \gamma(\phi) = j_s(\phi)/j_{c1}=\sin(\phi) + g\sin(2\phi)
  . \label{Eq:CPR(g)}
\end{equation}
Looking for an extremum of $\gamma(\phi)$ we find that this extremum is reached for the phase $\phi_0$ such that 
\begin{equation}
  \cos(\phi_0) = \frac{-1\pm\sqrt{1+32g^2}}{8g}
  . \label{Eq:cos(Phi_0)}
\end{equation}
Note, that the solution with the "$-$" sign in Eq.~(\ref{Eq:cos(Phi_0)}) only appears for $|g|\geq 1/2$. 

\begin{figure}[!htb]
  \includegraphics{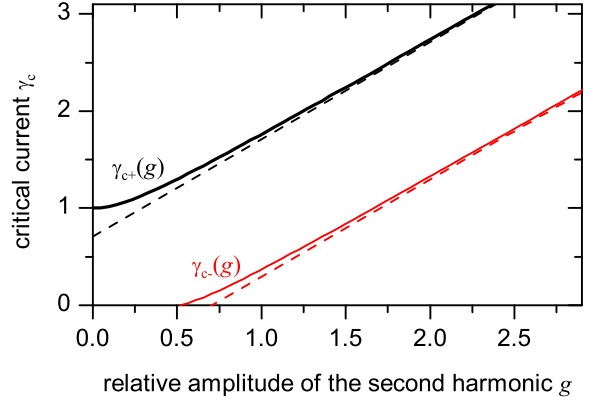}
  \caption{(Color online) 
    The dependences $\gamma_{c\pm}(g)$ and their asymptotic behavior for $g\to\infty$.
  }
  \label{Fig:Ic(g)}
\end{figure}

The primary (maximum) critical current is always given by substituting the solution (\ref{Eq:cos(Phi_0)}) with the "$+$" sign into Eq.~(\ref{Eq:CPR(g)})
\begin{equation}
  \gamma_{c+}(g) = \frac{1}{32|g|}
  \left( \sqrt{1+32g^2}+3 \right)^\frac32
  \left( \sqrt{1+32g^2}-1 \right)^\frac12
  . \label{Eq:gamma_c+}
\end{equation}
This dependence is shown in Fig.~\ref{Fig:Ic(g)}. For small $g$ it behaves as $\gamma_{c+}(g\to0)\approx 1 + 2g^2+O(g^4)$, while 
\begin{equation}
  \gamma_{c+}(g\to\pm\infty) \approx 
  \frac{1}{\sqrt2} \pm g \pm \frac{1}{16g}
  . \label{Eq:gamma_c+:large-g}
\end{equation}
For $|g|< 1/2$, the critical current is simply given by Eq.~(\ref{Eq:gamma_c+}). The physics is similar to a JJ with $g=0$.

For $|g| \geq 1/2$, the secondary critical current appears, see Fig.~\ref{Fig:Ic(g)}. It corresponds to $\phi_0$ with the ``$-$'' sign in Eq.~(\ref{Eq:cos(Phi_0)}) and is given by
\begin{equation}
  \gamma_{c-}(g) = \frac{1}{32|g|}
  \left( \sqrt{1+32g^2}-3 \right)^\frac32
  \left( \sqrt{1+32g^2}+1 \right)^\frac12
  . \label{Eq:gamma_c-}  
\end{equation}
Note, that $\gamma_{c-}(g\to\pm\infty)\approx \frac{-1}{\sqrt2} \pm g \pm \frac{1}{16g}$. It is interesting that the difference 
\begin{equation}
  \Delta\gamma_c=\gamma_{c+}(g)-\gamma_{c-}(g)\approx\sqrt{2}
  , \label{Eq:Delta-gamma_c}
\end{equation}
is almost constant as can be seen in Fig.~\ref{Fig:Ic(g)}. The largest deviation of $8\units{\%}$ takes place at $|g|=1/2$. From Eqs.~(\ref{Eq:gamma_c+}) and (\ref{Eq:gamma_c-}) one can see that $\gamma_{c\pm}(-g)=\gamma_{c\pm}(+g)$, therefore in Fig.~\ref{Fig:Ic(g)} we show $\gamma_{c\pm}(g)$ only for positive $g$.

\begin{figure*}[!htb]
  \includegraphics{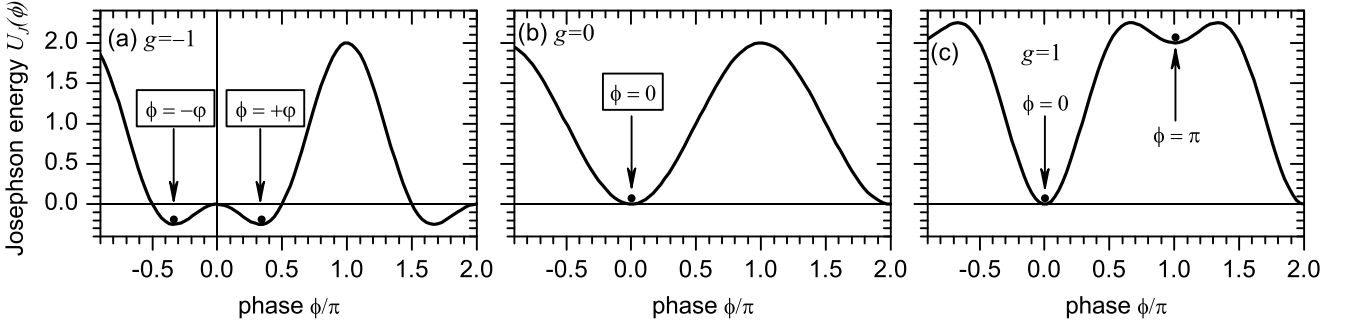}
  \caption{%
    Josephson energy $U_J$ of the system as a function of the phase $\phi$ for three cases: (a) $g=-1$, (b) $g=0$ and (c) $g=1$.
  }
  \label{Fig:U(phi)}
\end{figure*}

An important question is: can one observe $\gamma_{c-}$ in experiment? To answer this question we have to consider the case of negative and positive second harmonic separately.

1. For $g \geq 1/2$, and at $\gamma=0$ the system has two stable states $\phi=0$ (with lower energy) and $\phi=\pi$ (with higher energy), see Fig.~\ref{Fig:U(phi)}(c). The critical current $\gamma_{c-}$ corresponds to the ``depinning'' of the Josephson phase from the high energy $\pi$-state, while $\gamma_{c+}$ corresponds to a depinning from the low energy $0$-state. If initially the junction is in the 0-state, one just measures $\gamma_{c+}$ (\ref{Eq:gamma_c+}). On the other hand, if initially the JJ was in the $\pi$-state, at $\gamma_{c-}$ the Josephson phase starts moving. Depending on damping, it may either result in a stationary phase motion (for low damping) or the phase may go down to the neighboring lower energy $0$-state and stick there (for high damping). Thus, an underdamped JJ which was initially in the $\pi$-state will switch to the resistive state at $\gamma_{c-}$, while the overdamped JJ will just switch from the $\pi$-state to the $0$-state at $\gamma_{c-}$ and will switch to the resistive state only when the bias current is further increased above $\gamma_{c+}$. We note here that the probability to find the JJ initially in the $\pi$-state may not be very low. For example, for large $g$, the ratio of the energy difference between the $0$- and the $\pi$ states to the maximum barrier height goes like $[U(\pi)-U(0)]/U_\mathrm{max}\approx 1/g$, \ie, becomes negligible. This means that during switching from the voltage state to the Meissner state the phase may stick in the $\pi$-state with a probability close to $50\units{\%}$. If one is able to determine experimentally both $I_{c-}=I_{c1}\gamma_{c-}$ and $I_{c+}=I_{c1}\gamma_{c+}$, one will then be able to calculate $I_{c1}$, $I_{c2}$ and $g$ from experimental data. For large $g$ this calculation is straightforward:
\begin{eqnarray}
  I_{c1} &\stackrel{(\ref{Eq:Delta-gamma_c})}{\approx}&
  \frac{1}{\sqrt{2}}(I_{c+}-I_{c-})
  ; \label{Eq:I_c1}\\
  I_{c2} &\stackrel{(\ref{Eq:gamma_c+:large-g})}{\approx}&
  I_{c+}-\frac{I_{c1}}{\sqrt{2}}
  . \label{Eq:I_c2}
\end{eqnarray}

\begin{figure}[!htb]
  \includegraphics{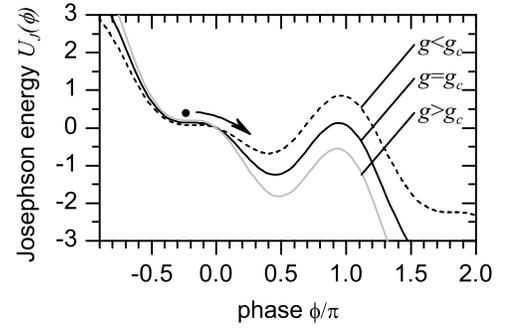}
  \caption{%
    The Josephson potential energy $U(\phi)$ at $\gamma=\gamma_{c-}(g)$ for $g<g_c$, $g=g_c\approx-1.26$ and $g>g_c$.
  }
  \label{Fig:U(phi)@g}
\end{figure}

2. For $g \leq -1/2$, the system has two degenerate stable states $\phi=\pm\varphi=\pm\arccos(-1/2g)$ ($\varphi$-junction), as shown in Fig.~\ref{Fig:U(phi)}(a). The critical current $\gamma_{c-}$ corresponds to the escape of the phase from the state $-\varphi$ towards the state $+\varphi$, while $\gamma_{c+}$ corresponds to the escape of the phase from the state $+\varphi$ towards $2\pi-\varphi$ over the large potential barrier. If initially the JJ is in the $+\varphi$ state, in experiment one observes only $\gamma_{c+}$. If initially the system is in the $-\varphi$ state, then, upon exceeding $\gamma_{c-}$, the phase moves towards the $+\varphi$-state and either ends up being trapped there (typical for an overdamped JJ) or may continue moving further switching the JJ into the voltage state. Which of these possibilities is realized depends not only on damping but also on the height of the potential barrier which the phase should overcome to keep moving continuously. This barrier, in turn, depends on $g$, see Fig.~\ref{Fig:U(phi)@g}.

A numerical study shows that the JJ with vanishing damping will not be trapped in the $+\varphi$-state for $g<g_c\approx-1.26$. At this value of $g$ the height of the barrier is equal to or smaller than the initial energy of the phase-particle, as shown in Fig.~\ref{Fig:U(phi)@g}.

%\Q{Is analytics possible?}

Thus, for a sufficiently large and negative second harmonic, one can observe two critical currents on the $I$--$V$ characteristic (IVC). Actually, the typical way to trace the IVC by sweeping the current $I$ from $-I_\mathrm{max}$ to $+I_\mathrm{max}$ and then back to $-I_\mathrm{max}$ will most probably result only in $\pm I_{c-}$ to be visible. To see $+I_{c+}$ one should sweep from 0 to $+I_\mathrm{max}$, to $0$, to $+I_\mathrm{max}$, etc., i.e. only the positive part of the IVC. To see $-I_{c+}$ one should trace only the negative half of the IVC. Simulations of the IVCs using different sweep sequences confirmed this prediction.

\subsection{Plasma waves}

One of the unique properties of Josephson junctions is the presence of the plasma frequency $\omega_p$. In a short JJ, $\omega_p$ is the frequency of the eigenoscillations of the Josephson phase at zero applied bias current. In a LJJ this is true only if the phase is uniform. In general, electromagnetic waves can propagate along the LJJ only if their frequency is above $\omega_p$. The dispersion relation $\omega(k)=\sqrt{1+k^2}$ in a conventional LJJ has a gap from 0 to 1. In quantum circuits the plasma gap $\hbar\omega_p \gg k_BT$ protects the circuit from thermally excited plasmons. It also defines the attempt frequency during the thermal escape from the zero voltage state as well as it defines the energy level spacing in the quantum regime. Therefore it is important to look into the dispersion relation and the plasma gap in the LJJ with a second harmonic in the CPR.

\begin{figure}[!htb]
  \begin{center}\includegraphics{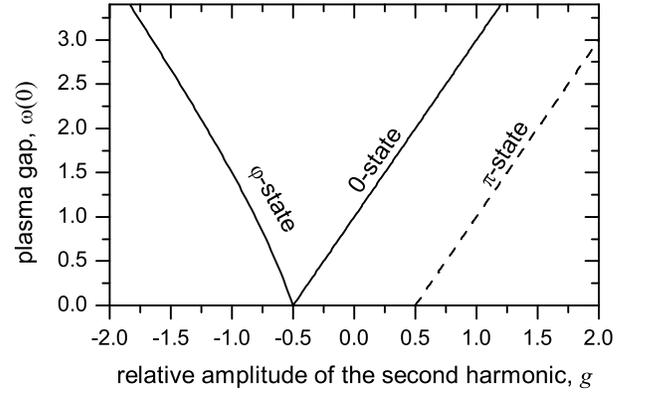}\end{center}
  \caption{
    The gap in the linear plasma wave spectrum as a function of $g$.
  }
  \label{Fig:gap(g)}
\end{figure}

The derivation of the dispersion relation $\omega(k)$ for plasma waves propagating along the junction is straightforward. We substitute the small amplitude wave solution $\phi=\phi_s(\gamma) + A\exp[i(kx-\omega t)]$ into Eq.~(\ref{Eq:sG}) without a damping term. $\phi_s(\gamma)$ is the uniform phase, which is a solution of the static Eq.~(\ref{Eq:sG}). When $\gamma=0$, it takes the value of $\phi_s=0$ (\ref{Eq:sol:0}) for $|g|\leq\frac12$, $\phi_s=0,\pi$  (\ref{Eq:sol:0},\ref{Eq:sol:pi}) for $g>\frac12$ and $\phi_s=\pm\varphi$  (\ref{Eq:sol:varphi}) for $g<-\frac12$. As a result one gets the dispersion relation 
\begin{equation}
  \omega(k) = \sqrt{\omega_p^2(g,\gamma) + k^2}
  , \label{Eq:omega(k)}
\end{equation}
where $\omega_p(g,\gamma)$ is the plasma gap, which depends on the ground state, the amplitude of the second harmonic $g$ and on the applied bias current $\gamma$. In a conventional ($g=0$) LJJ 
\begin{equation}
  \omega_p^0(0,\gamma)=\sqrt[4]{1-\gamma^2}
  , \label{Eq:ConvLJJ:omega_p(gamma)}
\end{equation}
where the superscript 0 stands for the ground state $\phi=0$. In the general case $\omega_p(g,\gamma)$ can be calculated as
\begin{equation}
  \omega_p(g,\gamma) = \sqrt{
    \cos\left[ \phi_s(\gamma) \right]+2g\cos\left[ 2\phi_s(\gamma) \right]
  }
  . %\label{Eq:}
\end{equation}
Since the static phase $\phi_s(\gamma)$ is a solution of the transcendential equation $\sin\phi_s+g\sin2\phi_s=\gamma$, one cannot write the explicit expressions for $\omega_p(g,\gamma)$ except for some limiting cases. When $\gamma=0$ one obtains
\begin{subequations}
  \begin{eqnarray}
    \omega_p^{0}(g,0)  &=& \sqrt{2g+1},    \mbox{ in 0-state, } g\geq -\frac12
    ; \label{Eq:omega_0}\\
    \omega_p^\pi(g,0)  &=& \sqrt{2g-1},\mbox{in $\pi$-state, } g\geq +\frac12
    ; \label{Eq:omega_pi}\\
    \omega_p^\varphi(g,0) &=&\sqrt{\frac{1}{2g} - 2g},
    \mbox{ in $\varphi$-state, } g<-\frac12
    . \label{Eq:omega_varphi}
  \end{eqnarray}
  \label{Eq:omega_p(g)}
\end{subequations}
The plot of the gap $\omega_p(g,0)$ is shown in Fig.~\ref{Fig:gap(g)}. The plasma gap closes and opens again at $g=-1/2$. The slope from the left is $-2\varepsilon$, while from the right it changes abruptly to $+4\varepsilon$, where $\epsilon=|g+\frac12|$. 

The dependence of the plasma gap $\omega_p(g,\gamma)$ on $\gamma$ looks qualitatively similar to Eq.~(\ref{Eq:ConvLJJ:omega_p(gamma)}), but with properly renormalized $\omega_p(g,0)$, see Eqs.~(\ref{Eq:omega_p(g)}), and $\gamma_c(g)$, see Eqs.~(\ref{Eq:gamma_c+}) and (\ref{Eq:gamma_c-}).

For the weakly biased $\gamma\ll1$ state corresponding to the ground state $\phi=0$ one finds
\begin{equation}
  \omega_p^0(g,\gamma)\approx 
  \sqrt{2g+1}-\frac{8g+1}{4(2g+1)^{5/4}}\gamma^2
  ,\quad \gamma\to0
  . \label{Eq:omega_p^0(g,gamma0)}
\end{equation}
For $\gamma\to\gamma_{c+}$, such that $\delta\gamma=\gamma_{c+}-\gamma\ll1$ the result is
\begin{equation}
  \omega_p^0(g,\gamma)\approx 
  \left[ \frac{(q+3)(q-1)q^2}{16g^2} \right]^{1/8}\sqrt[4]{\delta\gamma}
  , %\label{Eq:}
\end{equation}
where we have introduced the new variable
\begin{equation}
  q = \sqrt{1+32g^2}
  , \label{Eq:q}
\end{equation}
which is also used below to make formulas more compact.

For the ground state $\phi_s=\pi$, the weakly biased state will have the plasma frequency

\begin{equation}
  \omega_p^\pi(g,\gamma)\approx
  \sqrt{2g-1}-\frac{8g-1}{4(2g-1)^{5/2}}\gamma^2
  , %\label{Eq:}
\end{equation}
while in the pre-critical region $\delta\gamma=\gamma_{c-}(g)-\gamma\ll1$
\begin{equation}
  \omega_p^\pi(g,\gamma)\approx 
  \left[ \frac{(q-3)(q+1)q^2}{16g^2} \right]^{1/8}\sqrt[4]{\delta\gamma}
  . %\label{Eq:}
\end{equation}
\begin{figure}[!htb]
  \includegraphics{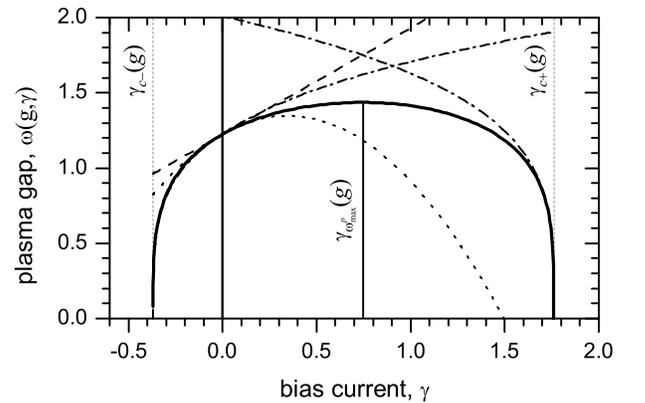}
  \caption{%
    $\omega_p^\varphi(g,\gamma)$ as a function of $\gamma$ for $g=-1$ and the ground state $\phi=+\varphi$ (thick solid line). Other lines show the approximation of this dependence in the vicinity of $-\gamma_{c-}$ (dash dotted line), $\gamma=0$ (dashed line)
  }
  \label{Fig:omega_p^varphi(gamma)}
\end{figure}

A much more interesting behavior of $\omega_p^\varphi(g,\gamma)$ is observed for $\varphi$-junctions. The behavior of $\omega_p^\varphi(g,\gamma)$ as a function of $\gamma$ is shown in Fig.~\ref{Fig:omega_p^varphi(gamma)}. One can see that $\omega_p^\varphi(g,\gamma)$ has a maximum not at $\gamma=0$, but at 
\begin{equation}
  \gamma_{\omega_p^\mathrm{max}}(g)=\pm\frac34\sqrt{1-\frac{1}{64g^2}}.
  \label{Eq:gamma_omega_max(g)}
\end{equation}
for the $\pm\varphi$ state, accordingly.

This feature is a direct consequence of the asymmetry of the potential well in which the phase-particle is trapped in the ground state. For comparison, this well is symmetric for 0 and $\pi$ ground states.

The behaviour $\omega_p^\varphi(g,\gamma)$ at $\gamma\to0$ can be approximated as 
\begin{equation}
  \omega_p^\varphi(g,\gamma)\approx 
  \sqrt{\frac{1-4g^2}{2g}} + \frac32 \sqrt{\frac{2g}{1-4g^2}}\gamma
  -\frac18 \sqrt{\frac{(2g)^3}{(1-4g^2)^5}}(32g^2+13)\gamma^2
  . %\label{Eq:gamma_p^varphi(gamma)}
\end{equation}

For $\gamma$ in the vicinity of $-\gamma_{c-}$, we define $\delta\gamma=\gamma+\gamma_{c-}$ and obtain
\begin{equation}
  \omega_p^\varphi(g,\gamma)\approx 
  \left[ \frac{(q-3)(q+1)q^2}{16g^2} \right]^{1/8}\sqrt[4]{\delta\gamma}
  . %\label{Eq:}
\end{equation}
For $\gamma$ in the vicinity of $\gamma_{c+}$, we difine $\delta\gamma=\gamma_{c+}-\gamma$ and obtain
\begin{equation}
  \omega_p^\varphi(g,\gamma)\approx 
  \left[ \frac{(q+3)(q-1)q^2}{16g^2} \right]^{1/8}\sqrt[4]{\delta\gamma}
  . %\label{Eq:}
\end{equation}

\subsection{Josephson Vortices}

\begin{figure*}[!htb]
  \begin{center}
    \includegraphics{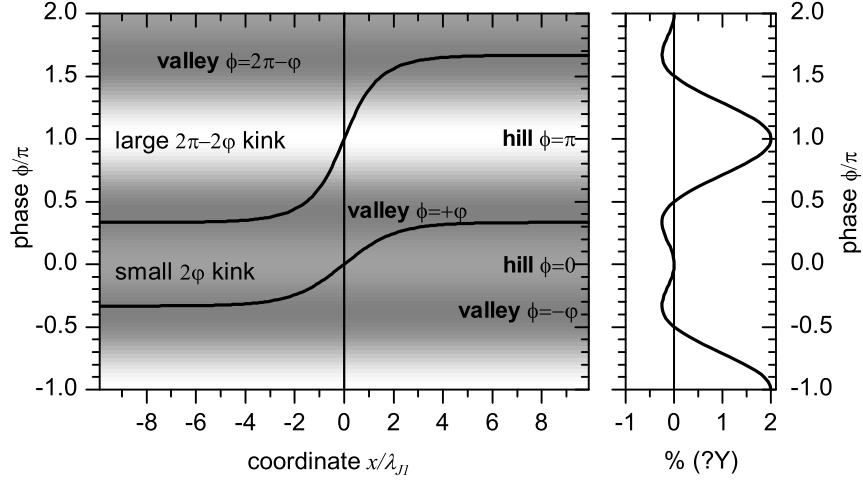}
  \end{center}
  \caption{%
    The shape of the small and the large fractional vortices (\ref{Eq:FractVort}). The Josephson energy profile $U(\phi)$ is shown in the right plot and also as a grayscale background in the left plot.
  }
  \label{Fig:FractVort}
\end{figure*}

The Josephson vortices which may exist in LJJs with a second harmonic can be very different from Josephson vortices in conventional LJJs. For the case $|g|<\frac12$ they still resemble usual sine-Gordon kinks. For $g<-\frac12$ there may be two kinds of solitons in the system: the smaller one $\phi_1(x)$carrying the flux $\Phi_1=\Phi_0\varphi/\pi$ ($2\varphi$ vortex) and a bigger one $\phi_2(x)$ carrying the flux $\Phi_2=\Phi_0-\Phi_1=\Phi_0(1-\varphi/\pi)$ ($2\pi-2\varphi$ vortex), see Refs.~\cite{Mints:1998:SelfGenFlux@AltJc,Mints:2000:SelfGenFlux@GB,Mints:2001:FracVortices@GB,Mints:2002:NonLocal+FracVortices,Buzdin:2003:phi-LJJ} for details. The shape of these vortices is given by the following formulas
\begin{subequations}
  \begin{eqnarray}
    \phi_1(x) &=& 2 \arctan \left[ 
      \tanh\left( \frac{x\sin\varphi}{2} 
      \tan\left( \frac{\varphi}{2} \right)\right) 
    \right]
    ; \label{Eq:FractVortSmall}\\
    \phi_2(x) &=& \pi+2 \arctan \left[ 
      \tanh\left( \frac{x\sin\varphi}{2} 
      \cot\left( \frac{\varphi}{2} \right)\right) 
    \right]
    , \label{Eq:FractVortLarge}
  \end{eqnarray}
  \label{Eq:FractVort}
\end{subequations}
and is shown in Fig.~\ref{Fig:FractVort}.

Such kinks were extensively studied in the framework of the double sine-Gordon equation and have some very interesting properties, \eg, one of them has an eigen oscillation frequency\cite{Kumar:1988:Solitons:InternalModes,Hudak:1981:dsG:IntModes}. 

Here, we will focus our attention on the point $g=-\frac12$ where the transition from 0 to $\varphi$-junction takes place. Let us write a static version of Eq.~(\ref{Eq:sG}), \ie, the analog of a Ferrell-Prange equation ($g=-\frac12$)
\begin{equation}
  \phi_{xx} = 4\sin^3\frac{\phi}{2}\cos\frac{\phi}{2}
  . \label{Eq:FP@g-1/2}
\end{equation}
From here, by multiplying with $\phi_x$ and integrating we obtain
\begin{equation}
  \left( \fracd{\phi}{x} \right)^2 = -2\cos\phi-\sin^2\phi+C
  , \label{Eq:phi_x}
\end{equation}
where $C$ is an integration constant. To satisfy the boundary conditions at $x\to\pm\infty$ (vanishing magnetic field $\phi_x$) one must set $C=2$, so that the \rhs of Eq.~(\ref{Eq:phi_x}) becomes equal to $4\sin^4\frac{\phi}{2}$, \ie,   
\begin{equation}
  \left( \fracd{\phi}{x} \right) = \pm 2\sin^2\frac{\phi}{2}
  , \label{Eq:phi_x:1}
\end{equation}
Integrating we get
\begin{equation}
  \phi(x) = \pm2\arctan\frac{1}{|x|} + 2 \pi n,
  , \label{Eq:NonExpKink}
\end{equation}

where the $\pm$ sign corresponds to solitons of positive and negative polarity. Note that if one takes the $+$ sign and $n=0$ for $x<0$, then one should take the $-$ sign and $n=1$ for $x>0$ in Eq.~(\ref{Eq:NonExpKink}). One can see that the soliton tail is non-exponential, namely $\phi\approx1/x$ for $x\to\pm\infty$.

\subsection{Penetration of magnetic field}

Let us consider the penetration of magnetic field into a semi-infinite LJJ ($x=0\ldots\infty$). Note that $\lambda_{J1}$ (\ref{Eq:lambda_J1}) just defines the unit length. It does not have the sense of the magnetic field penetration depth anymore (see below). 

The phase distribution over the junction is determined by the static version of Eq.~(\ref{Eq:sG})
\begin{equation}
  \phi_{xx} = \sin \phi + g \sin(2\phi)
  \label{Eq:FP}
\end{equation}

Below we consider the penetration of magnetic field for three different ground states of the system.

\paragraph{The ground state $\phi(\infty) =0$,} which is realized for $g>-1/2$. In this case the first integral of Eq.~(\ref{Eq:FP}) is
\begin{equation}
  \frac{1}{2}\phi_x^2 = 
  (1-\cos\phi) + \frac{g}{2} (1-\cos 2\phi)
  . \label{Eq:1stInt:0}
\end{equation}
The magnetic field penetration depth can be defined as 
\begin{equation}
  \lambda_H
  = \frac{1}{H(0)} \int_0^\infty H(x)\,dx
  = \frac{\Phi_{0}[\phi(\infty)-\phi(0)]}{2\pi \Lambda H(0)}
  = \lambda_{J1} \frac{[\phi(\infty)-\phi(0)]}{h(0)},
  \label{Eq:ScreenLength}
\end{equation}
where the dimensionless field is defined as usual,  $h(x)=\phi_x(x)=H(x) 2\pi \Lambda\lambda_{J1}/\Phi _{0}$, and 
\begin{equation}
  \Lambda = d_I 
  + \lambda_1 \tanh\left( \frac{d_1}{2\lambda_1} \right)
  + \lambda_2 \tanh\left( \frac{d_2}{2\lambda_2} \right)
  , \label{Eq:Lambda}
\end{equation}
is the effective magnetic thickness.

The boundary condition is $h(0)=h$. From Eq.~(\ref{Eq:1stInt:0}) we get
\begin{equation}
  8gy^2 - 4(2g+1)y + h^2=0
  , \label{Eq:h-y}
\end{equation}
where we defined $y=\sin^{2}\left[ \phi(0)/2 \right]$. Solving for $y$ we get
\begin{equation}
  y = \frac{(2g+1) - \sqrt{(2g+1)^2-2gh^2}}{4g}
  , \label{Eq:y}
\end{equation}
Substituting $\phi(0)=-2\arcsin(\sqrt{y})$ into Eq.~(\ref{Eq:ScreenLength}), we find
\begin{equation}
  \lambda_H^0 =\frac{\lambda _{J1}}{\sqrt{2g+1}},\quad h \ll 1
  ,\label{Eq:lambda_0}
\end{equation}
where the superscript 0 stands for the ground state $\phi=0$. At $g=-\frac12$ the screening length diverges (screening is non exponential) and 
$\lambda_H^0 = \lambda_{J1} \sqrt{\frac{2}{h}}$ for our case of small field  $h\ll1$.

Let us also find the \textbf{penetration field} $h_p$, i.e. the field at which vortices start penetrating into the LJJ. From Eq.~(\ref{Eq:h-y})
\begin{equation}
  h^{2}(y) = -8gy^2 + 4(2g+1)y.
\end{equation}
We should find the maximum value of $h$ under the constraints that $0\leq y \leq 1$. For $|g|<1/2$, the maximum value $h_p^0=2$ is reached for $y=1$. For $g>1/2$ the maximum value $h_p^0=(2g+1)/\sqrt{2g}$ is reached for $y=(2g+1)/4g$.

\paragraph{The state $\phi(\infty) =\pi$,} which is realized for $g>1/2$. In this case the first integral of Eq.~(\ref{Eq:FP}) is
\begin{equation}
  \frac{1}{2}\phi_x^2 = 
  (-1-\cos\phi) + \frac{g}{2} (1-\cos 2\phi)
  . \label{Eq:1stInt:pi}
\end{equation}
Following the same procedure we arrive at
\begin{equation}
  \lambda_H^\pi =\frac{\lambda _{J1}}{\sqrt{2g-1}},\quad h \ll 1
  . \label{Eq:lambda_pi}
\end{equation}
Note, that $\lambda_H^\pi$ diverges when $g\to1/2+0$.

To find the \textbf{penetration field} we have to search for the maximum of 
\begin{equation}
  h^2(y) = -8gy^2 +4(2g+1)y - 4
  , %\label{Eq:}
\end{equation}
with the constraints that $0\leq y \leq 1$. For $g \geq 1/2$ the maximum is reached at $y=(2g+1)/4g$ and is equal to $h_p^\pi=(2g-1)/\sqrt{2g}$.

\paragraph{The ground state $\phi=\pm\varphi$,} which corresponds to the domain $g<-1/2$. The first integral reads

\begin{equation}
  \frac12\phi_x^2 = 
  \frac{\left( \cos \varphi-\cos \phi \right)^2}{2\cos\varphi}.
  \label{Eq:1stInt:varphi}
\end{equation}
At $x=0$, $\phi_x=h$ and following the same procedure as before we get
\begin{equation}
  \lambda_H^\varphi = 
  \lambda _{J1}\frac{\sqrt{\cos\varphi}}{\sin\varphi}
  = \lambda _{J1}\sqrt{\frac{-2g}{4g^2-1}}
  . \label{Eq:lambda_varphi}
\end{equation}
For $g\to -1/2-0$, $\lambda_H^\varphi \to \lambda _{J1}\frac{1}{\sqrt{-2( 2g+1)}}$ in accordance with the previous result\cite{Buzdin:2003:phi-LJJ}.

To calculate the \textbf{penetration field} $h_p^\varphi$ we note that the LJJ may be in one of the ground states $\pm\varphi$. Without loosing generality we assume that it is $+\varphi$. We also assume that the LJJ is very long, but still has \emph{two edges} at $x=\pm L/2$. 
From Eq.~(\ref{Eq:1stInt:varphi}), at the left edge of the LJJ ($x=-L/2$) we have
\begin{equation}
  h=\phi_x=\frac{(\cos\phi -\cos\varphi)}{\sqrt{\cos\varphi}}
  . %\label{Eq:}
\end{equation}
and the maximum field (penetration field) $h_{p1}^\varphi$ corresponds to the penetration of the  $2\varphi$-soliton (\ref{Eq:FractVortSmall}) into the LJJ from the left edge. 
\begin{equation}
  h_{p1}^\varphi = \frac{(1-\cos\varphi)}{\sqrt{\cos\varphi}}
  = -\frac{2g+1}{\sqrt{-2g}}
  , \label{Eq:hL}
\end{equation}

Similarly, at the right edge $x=+L/2$ we have 
\begin{equation}
  h=\phi_x=\frac{(\cos\varphi-\cos\phi)}{\sqrt{\cos \varphi}}
  , %\label{Eq:h_p}
\end{equation}
and the maximum field (penetration field) $h_{p2}^\varphi$ corresponds to the penetration of the  $(2\pi-2\varphi)$-soliton (\ref{Eq:FractVortLarge}) into the LJJ from the right edge. 
\begin{equation}
  h_{p2}^\varphi = \frac{(\cos\varphi+1)}{\sqrt{\cos\varphi}}
  = \frac{1-2g}{\sqrt{-2g}}
  . \label{Eq:hR}
\end{equation}

Since $h_{p1}^\varphi<h_{p2}^\varphi$, the penetration of the flux in terms of vortices into the junction is not symmetric. At $h<h_{p1}^\varphi$ there are no vortices inside the LJJ, but the tails of two different kinks  penetrate into the LJJ from the edges: a tail of a $2\varphi$-kink from the left edge and the tail of a $(2\pi-2\varphi)$-kink from the right edge. At $h_{p1}^\varphi<h<h_{p2}^\varphi$, a single $2\varphi$-kink enters the LJJ from the left side and the phase behind the kink (between the kink and the left edge) sets to $-\varphi$. Thus, the next kink that should enter into the junction from the left side is a large $(2\pi-2\varphi)$-kink. But the field $h$ is still below $h_{p2}^\varphi$ --- the penetration field for the big kink. Thus, upon exceeding $h_{p1}^\varphi$, a single small vortex enters the LJJ and the field does not penetrate further until the field exceeds $h_{p2}^\varphi$. Upon exceeding $h_{p2}^\varphi$, a chain of alternating big, small, big, small, etc vortices enter into the LJJ from the left and from the right ends until they fill the whole LJJ with a certain density. 

\begin{figure*}[!htb]
  \includegraphics{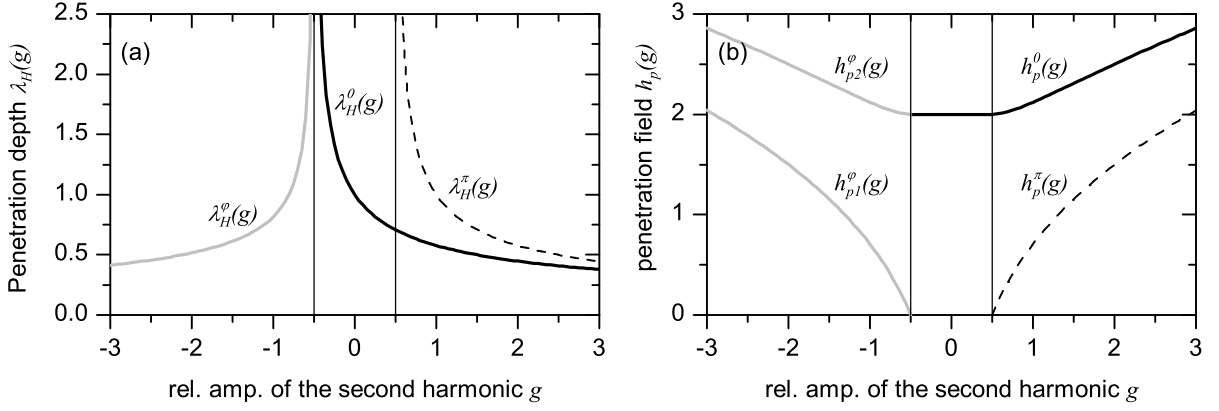}
  \caption{%
    Dependence of (a) the small magnetic field penetration depth $\lambda_H(g)$ and (b) the vortex penetration field $h_p(g)$ on $g$.
  }
  \label{Fig:Penetr(g)}
\end{figure*}

Fig.~\ref{Fig:Penetr(g)} visually summarizes the results on $\lambda_H(g)$ and $h_p(g)$ obtained above.

%\Q{How to see this state with a single small vortex? For example, inject it, apply bias current and try to see ZFS. If one succeeds to see such a ZFS, it will be a direct evidence of vortex charge!}

\subsection{Dependence $I_c(H)$}

The calculation of the $I_c(H)$ dependence for a short JJ of the length $2a=L$ and the width $w$ follows the usual procedure. We assume that the phase is linear with magnetic field,
\begin{equation}
  \phi=hx+\phi_{0}
  , \label{Eq:phase-ansatz}
\end{equation}
where $h=2H/H_{c1}$ is the dimensionless applied magnetic field. It is normalized to $H_{c1}=\Phi_0/\pi\lambda_{J1}\Lambda$ which is the penetration field of the JJ with $g=0$. 

%$\phi_{1,2}$ is the arbitrary constant phase, which is a free parameter. Subscripts refer to two different solutions, see below, and at this point should be considered as peculiar notation and do not have any special meaning.

The total supercurrent through the junction normalized to the maximum supercurrent of the first harmonic $j_{c1}2aw\lambda_J^2$ is given by
\begin{eqnarray}
  &&i_s(h,g,\phi_0) =
  \frac{1}{2a}\int_{-a}^{+a} 
  \left[ \sin(hx+\phi_0)+g\sin(2hx+2\phi_0) \right]\,dx
  \nonumber\\
  &&= \frac{2\sin(ha)\sin(\phi_0)+g\sin(2ha)\sin(2\phi_0)}{2ha}
  . \label{Eq:I}
\end{eqnarray}
Note, that $ha=\pi\Phi/\Phi_0$ is a normalized flux inside the junction.
To find the maximum value of the supercurrent we should find the maximum of $I_s(h,\phi_0)$ with respect to $\phi_0$. The maximum is reached when $\partial i_s/\partial\phi_0=0$. This yields the following equation for $\cos(\phi_0)$:
\begin{equation}
  8g\sin(2ha) \cos^2(\phi_0)
  +4\sin(ha) \cos(\phi_0)
  -4g\sin(2ha)=0
  . \label{Eq:Eq4phi_0}
\end{equation}
This equation has two solutions, which we denote as $\cos(\phi_1)$ and $\cos(\phi_2)$:
\begin{eqnarray}
  \cos(\phi_1) &=& \frac{-1+\sqrt{1+32g^2\cos^2(ha)}}{8g\cos(ha)}
  ; \label{Eq:cos-phi_1}\\
  \cos(\phi_2) &=& \frac{-1-\sqrt{1+32g^2\cos^2(ha)}}{8g\cos(ha)}
  . \label{Eq:cos-phi_2}
\end{eqnarray}
To get $I_c(H)$ we substitute $\phi_{1,2}$ from Eqs.~(\ref{Eq:cos-phi_1}) and (\ref{Eq:cos-phi_2}) into Eq.~(\ref{Eq:I}), i.e., $I_c(h,g)=I(h,g,\phi_{1,2}(h,g))$. In principle, we get two branches of $I_c(h)$. 

\begin{figure}[!htb]
  \includegraphics{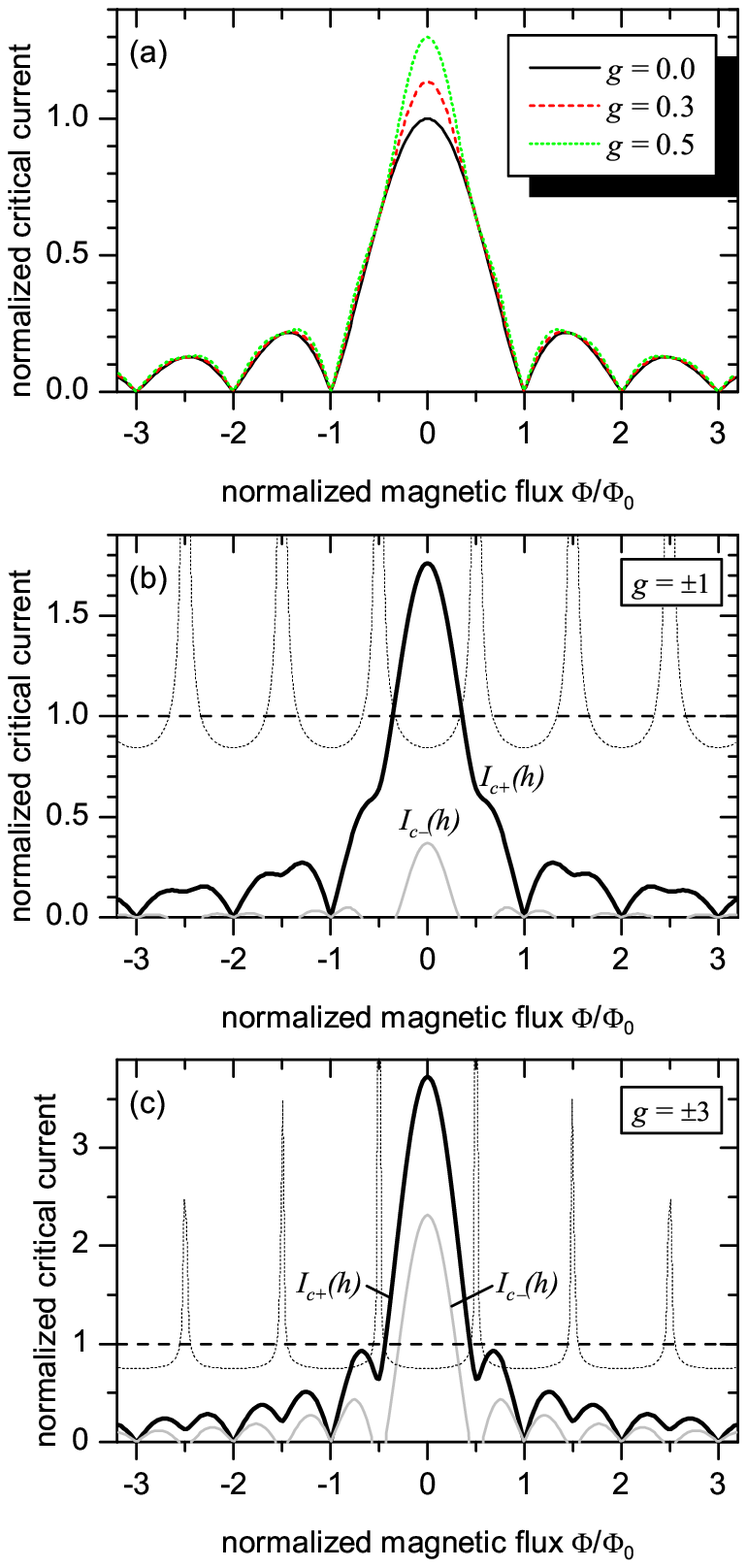}
  \caption{(Color online)
    The dependences $I_{c+}(h)$ and $I_{c-}(h)$ calculated using Eq.~(\ref{Eq:I}) with $\phi_{1,2}$ given by Eqs.~(\ref{Eq:cos-phi_1}) and (\ref{Eq:cos-phi_2}). 
    (a) for $g=0$, $\pm0.3$, $\pm0.5$ only the main branch $I_{c+}(h)$ is present;
    (b) for $g=\pm1$;
    (c) for $g=\pm3$. 
    In (b) and (c) the r.h.s. of (\ref{Eq:cos-phi_2}) are shown by the dotted line. The solution corresponding to $I_{c-}(h)$ exists only when this line goes below 1.
  }
  \label{Fig:Ic(H)@g}
\end{figure}

The solution given by Eq.~(\ref{Eq:cos-phi_1}) always has $|\cos(\phi_1)|\leq1$ and, upon substitution into Eq.~(\ref{Eq:I}), yields the upper branch $I_{c+}(H)$, which corresponds to $I_c(H)$ in a conventional JJ with $g=0$. Instead, the second solution given by Eq.~(\ref{Eq:cos-phi_2}) has $|\cos(\phi_2)|>1$ if $|g|<1/2$, i.e. it is irrelevant for $|g|<1/2$. For $|g|>1/2$, $|\cos(\phi_2)|$ may or may not be below one, depending on the value of the applied field, and, for some ranges of magnetic field, yields $I_{c-}(H)$. The calculated $I_{c\pm}(h)$ plots are shown in Fig.~\ref{Fig:Ic(H)@g} for different values of $g$. One can see that features with twice shorter period in $H$ appear as $|g|$ increases. In the limit $|g|\to\infty$, $I_{c+}(H)$ again becomes a Fraunhofer pattern, but with twice shorter period.

We note that $I_{c}(h,g)$ looks exactly the same as $I_{c}(h,-g)$. Indeed, changing $+g$ to $-g$ changes the sign of $\cos(\phi_{1,2})$ in Eqs.~(\ref{Eq:cos-phi_1}) and (\ref{Eq:cos-phi_2}). This corresponds to the change from $\phi_{1,2}$ to $\pi-\phi_{1,2}$. Looking at Eq.~(\ref{Eq:I}) one sees that $\sin(\phi_0)$ is a symmetric function of $\phi_0$ with respect to $\phi_0=\pi/2$, \ie, does not change, while $g\sin(2\phi_{1,2})=-g\sin(2(\pi-\phi_{1,2}))$ i.e. $I_s(h,g,\phi_{1,2})=I_s(h,-g,\pi-\phi_{1,2})$. Thus, $I_{c}(h,g)=I_{c}(h,-g)$. From a practical point of view, this means that \emph{one cannot distinguish between a JJ with positive and negative second harmonic by measuring $I_c(H)$ in a small JJ}. 

For $|g|<1/2$, see Fig.~\ref{Fig:Ic(H)@g}(a), the plots $I_{c+}(h)$ look very similar to the usual Fraunhofer dependence except for the region around $h=0$. In practice, one can detect the presence of a weak second harmonic by measuring $I_c(H)$ and calculating the ratio of $I_c(H_1)$ at the first side maximum to the $I_c(0)$. In a conventional JJ this ratio is $\approx 2/3\pi\approx0.21$. In the presence of the second harmonic $I_c(H_1)$ stays almost the same, while $I_c(0)$ changes with $g$, see Eq.~(\ref{Eq:gamma_c+}), reaching $3\sqrt{3}/4\approx1.3$ at $|g|\to 1/2$. Thus $I_c(H_1)/I_c(0)$ changes from 0.21 to $8\sqrt{3}/27\pi\approx0.16$ for $|g|$ from 0 to $1/2$.

For  $|g|>1/2$, see Fig.~\ref{Fig:Ic(H)@g}(b)--(c), the second lower branch $I_{c-}(h)$ can be seen. Note that this branch appears not for all values of $h$, but only for those where the \rhs of Eq.~(\ref{Eq:cos-phi_2}) $\leq 1$. 
The values of the \rhs of Eq.~(\ref{Eq:cos-phi_2}) are shown in Fig.~\ref{Fig:Ic(H)@g}(b)--(c) by the dashed line. Note, that $I_{c-}(h)<I_{c+}(h)$ for all values of $h$. 

%\Q{Simulations of $I_c(H)$ in a short and long JJ???}

\subsection{Shapiro steps}

A junction with a second harmonic shows not only integer, but also semi-integer Shapiro steps when a microwave voltage is applied, even when the capacitance of the JJ $C=0$. Here we consider a simple model of applied dc + ac voltage
\begin{equation}
  V(t) = V_0 + V_1 \cos(\omega t)
  . \label{Eq:V(t)}
\end{equation}
From the second Josephson relation $\phi_t=2\pi V/\Phi_0$, the phase in the junction will have the form
\begin{equation}
  \phi(t) = \phi_0 + \omega_J t + A \sin(\omega t)
  , \label{Eq:phi(t)}
\end{equation}
where $\phi_0$ is an arbitrary phase shift (integration constant), $\omega_J=2\pi V_0/\Phi_0$ is the Josephson frequency (frequency of internal oscillations in the JJ), and $A=V_1/\omega$ is the amplitude of the phase oscillations.

From here on we follow the standard procedure \cite{Barone:JosephsonEffect}. We substitute the phase ansatz (\ref{Eq:phi(t)}) into the expression for the supercurrent (\ref{Eq:CPR(g)}), expand terms like $\cos(A\sin(\omega t))$ and $\sin(A\sin(\omega t))$ in terms of Bessel functions, and find the dc component e.g. by averaging over time. Below we present the results for $\omega_J=n\omega$, where $n$ is an integer or half-integer number. 
\iffalse
\begin{figure*}[!htb]
  \includegraphics{Shapiro-I0}
  \includegraphics{Shapiro-I1}
  \caption{(Color online)
    $\gamma_0^\mathrm{max}(A)$ (left) and $\gamma_1^\mathrm{max}(A)$ (right) for several different values of $|g|$. Two solutions (marked as red and blue) are visible. For $g=0$ these two solutions correspond to the odd and even period of the usual $|\J_n(A)|$ dependence, but for $|g|>0$ they overlap for some $A$ and do not exist for other $A$. The values of $\cos(\phi_0)$ for both solutions are shown by corresponding red and blue dashed lines.
  }
  \label{Fig:ShapiroInt}
\end{figure*}
\fi

\textbf{For integer} $n$ the general formula for the supercurrent reads\cite{Kleiner1996:cAxisTunnelMicrowave}

\begin{equation}
  \overline{\gamma(A,\phi_0)}_{n} = \left|
    \sin(\phi_0)J_{n}(A) + g\sin(2\phi_0)J_{2n}(2A)
  \right|
  .\label{Eq:Shapiro:int}
\end{equation}
As we see the contribution of the Josephson supercurrent into the total dc current depends on $\phi_0$. The amplitude of the Shapiro step corresponds to the maximum of $\overline{\gamma(\phi)}_n$ with respect to $\phi_0$. The maximum is reached for 
\begin{equation}
  \cos(\phi_{0})=
  \frac{-\J_n(A)\pm\sqrt{\J_n^2(A)+32g^2\J_{2n}^2(A)}}{8g\J_{2n}(A)}
  . \label{Eq:Shapiro:int:cos_phi_0}
\end{equation}
Each of these two solutions is relevant only if $|\cos(\phi_0)|<1$. If, for given $A$, $|\cos(\phi_0)|<1$, we calculate $\phi_0$ as $\arccos$ of Eq.~(\ref{Eq:Shapiro:int:cos_phi_0}) and substitute it into Eq.~(\ref{Eq:Shapiro:int}). We get a rather bulky expression for $\gamma_n^\mathrm{max}(A)$, which we do not show here. Note, that taking $-\arccos()$ of Eq.~(\ref{Eq:Shapiro:int:cos_phi_0}) produces the same result. We also note that $\gamma_n^\mathrm{max}(A)$ does not change if we change $+g$ to $-g$. %The behavior of $\gamma_n^\mathrm{max}(A)$ for several different values of $|g|$ is shown in Fig.~\ref{Fig:ShapiroInt}.

\begin{figure}[!htb]
  \includegraphics{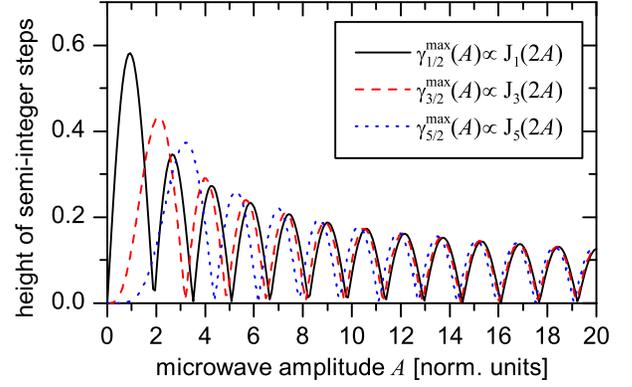}
  \caption{(Color online)
    The behavior of $\gamma_n^\mathrm{max}(A)$ for $n=\frac12,\frac32,\frac52$. 
  }
  \label{Fig:Shapiro:SemiInt}
\end{figure}

\textbf{For semi-integer} $n$, the supercurrent is given by\cite{Kleiner1996:cAxisTunnelMicrowave}
\begin{equation}
  \overline{\gamma(A,\phi_0)}_{n} =  
    g\sin(2\phi_0)J_{2n}(2A)
  . \label{Eq:Shapiro:semiint_n}  
\end{equation}
It is obvious that the maximum contribution to the dc current takes place for $\sin(2\phi_0)=\pm1$, therefore\cite{Kleiner1996:cAxisTunnelMicrowave}
\begin{equation}
  \gamma_n^\mathrm{max}(A) = g|J_{2n}(2A)|
  . \label{Eq:gamma-max-semiint}
\end{equation}
As we see, the semi-integer Shapiro steps appear as soon as $g\ne 0$. The behavior of $\gamma_n^\mathrm{max}(A)$ for $n=\frac12, \frac32$ is shown in Fig.~\ref{Fig:Shapiro:SemiInt} for $|g|=1$, but can be scaled to get the dependence for arbitrary $|g|$.

Recently, a formula for integer and semi-integer Shapiro steps which takes into account the capacitance of a JJ was obtained\cite{Kislinskii:2005:YBCO-Nb}. This formula is valid in the high frequency limit and, in principle, it allows to determine the sign of the second harmonic experimentally. On the other hand, this formula does not take into account the possibility of chaotic dynamics in the underdamped system, so its application may be problematic in some cases, and one should rely on numerical simultions.

\subsection{Zero Field Steps}

\begin{figure*}[!htb]
  \begin{center}
    \includegraphics{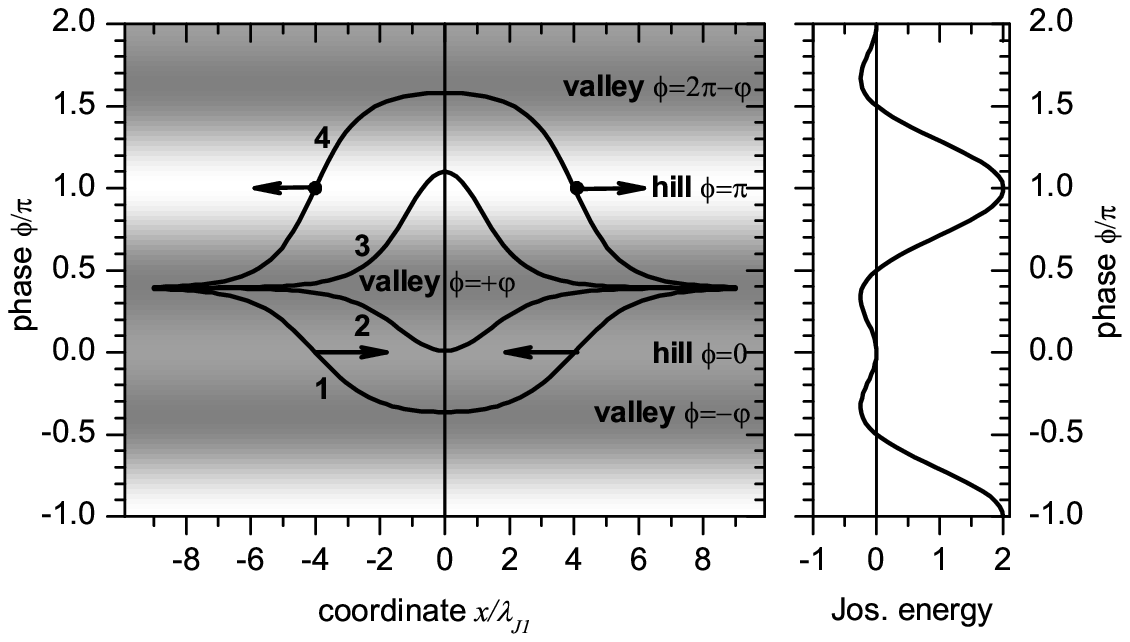}\\
  \end{center}
  \caption{%
    Reflection of a small fractional vortex $\Phi_1$ from the edge of the LJJ as a large fractional antivortex $-\Phi_2$. The junction is situated at $x<0$, the edge is at $x=0$. The reflection can be represented as a collision of a small vortex $+\Phi_1$ with a small antivortex $-\Phi_1$ at point $x=0$ in an infinite LJJ. They turn into a large antivortex $-\Phi_2$ and a large vortex $+\Phi_2$ after collision. Curves 1 to 4 show the evolution of the phase with time during this process. The Josephson energy profile $U(\phi)$ is shown in the right plot and also as a grayscale background in the left plot.
  }
  \label{Fig:Collision}
\end{figure*}
The presence of mobile fractional vortices for $g<-1/2$ allows the observation of half-integer zero field steps (ZFS) on the current-voltage characteristic similar to classical integer ZFS\cite{Fulton:ZFS,Levring:ZFS-ov-in,Pedersen:ZFS-cmp,Barone:JosephsonEffect}. The dynamics can be described as follows. A small vortex situated inside the junction subject to a driving force due to the bias current, moves along the junction. When it arrives at the edge $x=L$, the boundary condition requires that $\phi_x(L)=0$. As in the analysis of integer ZFS, we satisfy this boundary condition by considering a collision of a small $2\varphi$-vortex with its ``image'' --- a small $-2\varphi$-antivortex --- situated outside the LJJ. Thus, instead of treating the collision of a vortex with the boundary at $x=L$, we treat the collision of a vortex with an antivortex at $x=L$ in the absence of the boundary. Such a collision of a small vortex and an antivortex should inevitably result in the appearance of a large $(2\pi-2\varphi)$-vortex and a $(2\varphi-2\pi)$-antivortex after collision, see Fig.~\ref{Fig:Collision}. Thus, a $2\varphi$-vortex colliding with the boundary, reflects back as a $(2\varphi-2\pi)$-antivortex. Then the bias current pulls this large $(2\varphi-2\pi)$-antivortex towards the opposite boundary at $x=0$. There, it reflects as a small $2\varphi$-vortex, which is pulled towards the boundary $x=L$, and so on. The flux transfer per period is $\Phi_1-(-\Phi_2)=\Phi_0$, therefore the voltage across the junction is $V=\Phi_0 u/2L$. When the bias is icreased the velocity of vortices $u\to{\bar c}_{0}$ and the voltage approaches $V\to\Phi_0 {\bar c}_{0}/2L$ i.e. half of the usual ZFS. 

\begin{figure}[!htb]
  \begin{center}
    \includegraphics{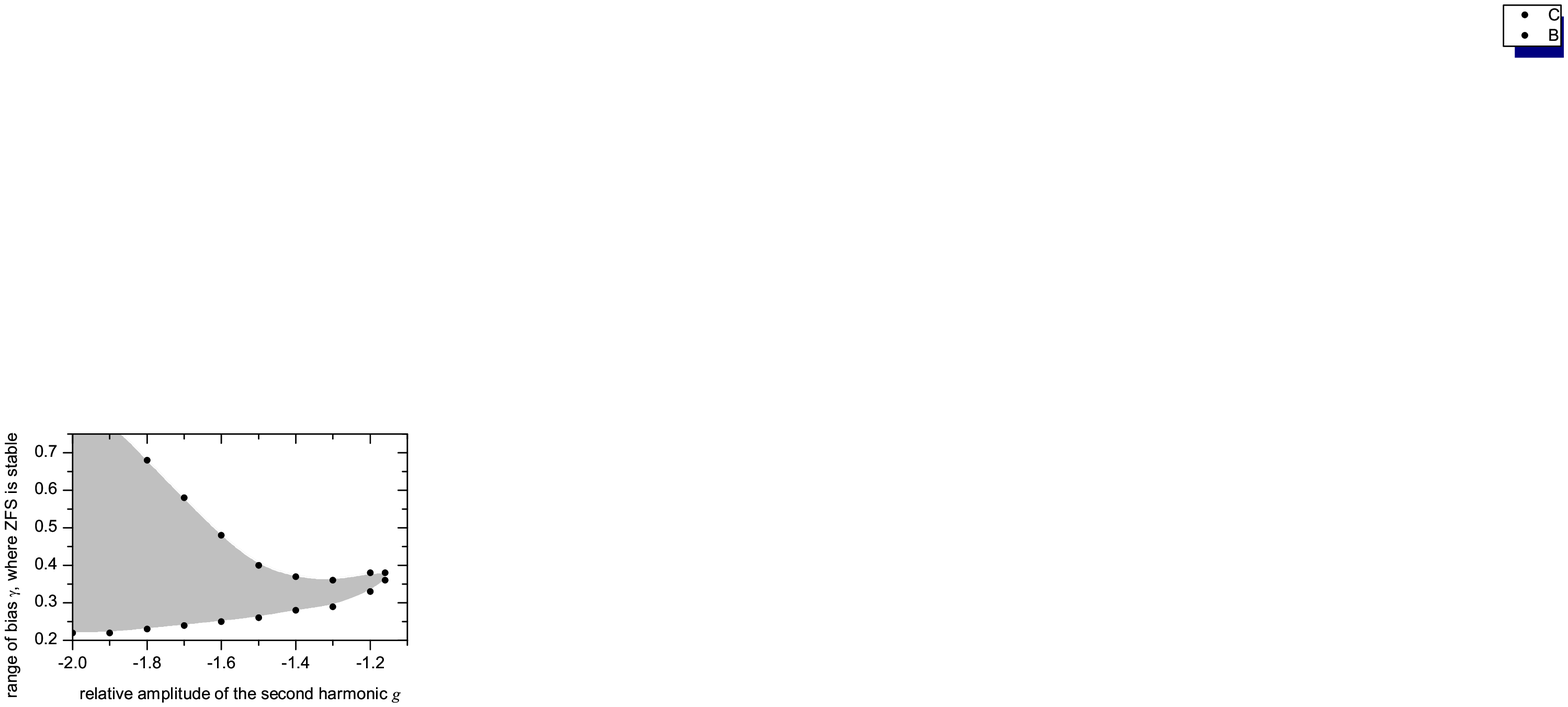}
  \end{center}
  \caption{%
    Numerically simulated range of stability (gray region) for half-integer ZFS as a function of $g$. Symbols show the values obtained numerically for a LJJ of length $L=20$ with $\alpha=0.1$. %Lines shown as a guide for the eyes.
  }
  \label{Fig:ZFS}
\end{figure}

However there is a problem. Even in the lossless case, the reflection of a small vortex as a large vortex requires an additional input of energy (which can be taken from kinetic energy or from the current source) to compensate for the energy difference between the small and the large vortices. Therefore, such a reflection may take place only when $\Phi_1\approx\Phi_2$, i.e. for rather large $|g|$. Indeed numerical simulations show that half integer ZFS can be observed for $g \lesssim -1$. The range of bias currents in which the first half-integer ZFS is stable is shown in Fig.~\ref{Fig:ZFS} for different values of $g$ and $\alpha=0.1$. On one hand, the value of $\alpha$ should be smaller than 1 to observe the dynamics. On the other hand, if $\alpha$ is too small, upon reflection one excites too many plasma waves that destroy the solitons and switch the junction into the normal state.

Note that the usual integer ZFS exist in such a LJJ also, and they are much more stable. The shape of the soliton is of course different from the classical sine-Gordon kink (having two maxima in $\phi_x(x)$, but its general behavior is the same.

\section{Conclusions}
\label{Sec:Conclusions}

We have investigated some properties of long Josephson junctions with an  arbitrarily strong amplitude $j_{c2}$ of the second harmonic in the current phase relation, which is always the case in the vicinity of the 0-$\pi$ transition where the amplitude of the first harmonic $j_{c1}$ vanishes. We have shown that in the case of low damping and $|g|>1/2$ ($g=j_{c2}/j_{c1}$) one can observe two critical currents and determine the amplitude of the first and the second harmonics in the current-phase relation experimentally.

For $|g|>1/2$ the second critical current may also be seen on the $I_c(H)$ dependence, while the dependence itself develops twice shorter periodicity in $H$. Unfortunately, one cannot deduce the sign of the second harmonic from the shape of the $I_c(H)$ dependence since it depends only on $|g|$. For $|g|<1/2$ the $I_c(H)$ dependence changes only slightly, but the presence of the second harmonic can be noticed by comparing the heights of the principal maximum and the first side maxima.

In the presence of the second harmonic, half integer Shapiro steps appear. Their amplitude is given by the simple formula (\ref{Eq:gamma-max-semiint}). The amplitude of the integer Shapiro steps depends on the rf amplitude in a complicated way. Unfortunately, one cannot deduce the sign of the second harmonic from the modulation of the Shapiro steps amplitude.

The most interesting property of the investigated system is that for $g<-1/2$ two types of fractional vortices may exist in the LJJ. One of them carries the flux $\Phi_1$ and the other $\Phi_2$ such that $\Phi_1+\Phi_2=\Phi_0$. The smaller vortex (carrying the flux  $\Phi_1<\Phi_2$) penetrates into the LJJ easier than the larger one, resulting in two different penetration fields $H_{p1}$ and $H_{p2}$. The presence of the fractional vortices and, therefore, of a strong \emph{negative} second harmonic, may be detected experimentally by the observation of half-integer zero field steps that should appear in LJJs with moderate damping $0.01<\alpha<1$.

\acknowledgments

We are grateful to A. Abdumalikov for reading the manuscript and for his suggestions. E.G. thanks the University of Bordeaux for hospitality and the ESF program PiShift for financial support. We also acknowledge the support by the DFG (project GO-1106/1) and by the Landesstiftung Baden-W\"urttemberg.

\bibliography{MyJJ,JJ,LJJ,SF,pi,SFS,software,ratch}

\end{document}